\documentclass[english,twocolumn,prl,amssymb,amsmath,superscriptaddress]{revtex4}
\usepackage[latin9]{inputenc}
\usepackage{graphicx,color,nicefrac}
\usepackage[hidelinks]{hyperref}
\hypersetup{colorlinks=false}
\usepackage{babel}
\usepackage{braket}
\usepackage{mathtools}
\usepackage[super]{nth}
\usepackage{textcomp}
\usepackage{siunitx}

\definecolor{mygreen}{rgb}{0,0.5,0}
\definecolor{myblue}{rgb}{0,0,0.75}
\definecolor{mymagenta}{cmyk}{0,1,0,0.12}
\definecolor{mygray}{rgb}{0.5,0.5,0.5}

\usepackage{umoline}

\begin{document}
\title{Environment-assisted quantum transport in a 10-qubit network}

\author{Christine Maier}
	\affiliation{Institute for Quantum Optics and Quantum Information,
	Austrian Academy of Sciences, Technikerstr. 21A, 6020 Innsbruck,
	Austria}
	\affiliation{Institute for Experimental Physics, University of Innsbruck,
	Technikerstr. 25, 6020 Innsbruck, Austria}
\author{Tiff Brydges}
	\affiliation{Institute for Quantum Optics and Quantum Information,
	Austrian Academy of Sciences, Technikerstr. 21A, 6020 Innsbruck,
	Austria}
	\affiliation{Institute for Experimental Physics, University of Innsbruck,
	Technikerstr. 25, 6020 Innsbruck, Austria}
\author{Petar Jurcevic}
	\affiliation{Institute for Quantum Optics and Quantum Information,
	Austrian Academy of Sciences, Technikerstr. 21A, 6020 Innsbruck,
	Austria}
	\affiliation{Institute for Experimental Physics, University of Innsbruck,
	Technikerstr. 25, 6020 Innsbruck, Austria}
\author{Nils Trautmann}
	\affiliation{Institute for Applied Physics,
	TU Darmstadt, 64289, Germany}	
\author{Cornelius Hempel}
	\affiliation{Institute for Quantum Optics and Quantum Information,
	Austrian Academy of Sciences, Technikerstr. 21A, 6020 Innsbruck,
	Austria}
    \affiliation{Institute for Experimental Physics, University of Innsbruck,
	Technikerstr. 25, 6020 Innsbruck, Austria}
	\affiliation{ARC Centre of Excellence for Engineered Quantum Systems, School of Physics, University of Sydney, NSW, 2006, Australia }	
\author{Ben P. Lanyon}
	\affiliation{Institute for Quantum Optics and Quantum Information,
	Austrian Academy of Sciences, Technikerstr. 21A, 6020 Innsbruck,
	Austria}
	\affiliation{Institute for Experimental Physics, University of Innsbruck,
	Technikerstr. 25, 6020 Innsbruck, Austria}
\author{Philipp Hauke}
	\affiliation{Kirchhoff-Institute for Physics, Heidelberg University, 69120 Heidelberg, Germany}
	\affiliation{Institute for Theoretical Physics, Heidelberg University, 69120 Heidelberg, Germany}
\author{Rainer Blatt}
	\affiliation{Institute for Quantum Optics and Quantum Information,
	Austrian Academy of Sciences, Technikerstr. 21A, 6020 Innsbruck,
	Austria}
	\affiliation{Institute for Experimental Physics, University of Innsbruck,
	Technikerstr. 25, 6020 Innsbruck, Austria}
\author{Christian F. Roos}
	\affiliation{Institute for Quantum Optics and Quantum Information,
	Austrian Academy of Sciences, Technikerstr. 21A, 6020 Innsbruck,
	Austria}
	\affiliation{Institute for Experimental Physics, University of Innsbruck,
	Technikerstr. 25, 6020 Innsbruck, Austria}
	\email{christian.roos@uibk.ac.at}
\date{\today}     

\begin{abstract}
The way in which energy is transported through an interacting system governs fundamental properties in many areas of physics, chemistry, and biology. 
Remarkably, environmental noise can enhance the transport, an effect known as environment-assisted quantum transport (ENAQT). 
In this paper, we study ENAQT in a network of coupled spins subject to engineered static disorder and temporally varying dephasing noise. The interacting spin network is realized in a chain of trapped atomic ions and energy transport is represented by the transfer of electronic excitation between ions.
With increasing noise strength, we observe a crossover from coherent dynamics and Anderson localization to ENAQT and finally a suppression of transport due to the quantum Zeno effect. 
We find that in the regime where ENAQT is most effective the transport is mainly diffusive, displaying coherences only at very short times. 
Further, we show that dephasing characterized by non-Markovian noise can maintain coherences longer than white noise dephasing, with a strong influence of the spectral structure on the transport efficiency. 
Our approach represents a controlled and scalable way to investigate quantum transport in many-body networks under static disorder and dynamic noise. 
\end{abstract}

\maketitle

\emph{Introduction.---}
\label{sec:Introduction}
The transport of energy through networks governs fundamental phenomena such as light harvesting in photosynthetic organisms~\cite{Blankenship:2002, Sension:2007, Caruso:2009} or properties of nano-fabricated quantum devices~\cite{Shohiro:2013, Lee:2005}.
Often, such systems are subject to static disorder, which for non-interacting particles suppresses transport through Anderson localization~\cite{anderson1958}. 
In realistic networks, coupling to environments such as phonon baths moreover induces dynamical noise that can lift Anderson localization, an effect known as environment-assisted quantum transport (ENAQT). This phenomenon has been postulated to be a key factor enabling the high efficiency of energy conversion in photosynthetic biomolecules \cite{Plenio2013,Lambert2013,Fassioli2013}. 
At large noise levels, the transport efficiency again decreases due to the quantum Zeno effect~\cite{Misra1977Zeno}. 
While the general phenomenology governing the transport efficiency is widely accepted, many works have been dedicated to understanding the influence of non-Markovian noise \cite{Thorwart:2009,Chen:2011,Chin:2013,Mohseni:2014,Jesenko:2013} as well as coherence \cite{Engel:2007,Lee:2007,Chin:2013,Rey:2013,Panitchayangkoon:2010,Mohseni:2008,Plenio:2008,Caruso:2009}. 
Here, engineered quantum systems provide a prime opportunity, by enabling controlled studies of energy transport under noisy environments. Recent experiments have started investigating the elementary building blocks of ENAQT, but were limited to at most four network nodes, represented by photonic wave-guides, classical electrical oscillators, superconducting qubits, or trapped ions~\cite{Viciani2015,Biggerstaff2016,Montiel2015,Mostame2012,Wallraff2018,Haeffner2018}. 

\begin{figure}[t!]
   \centering
   \includegraphics[scale=1]{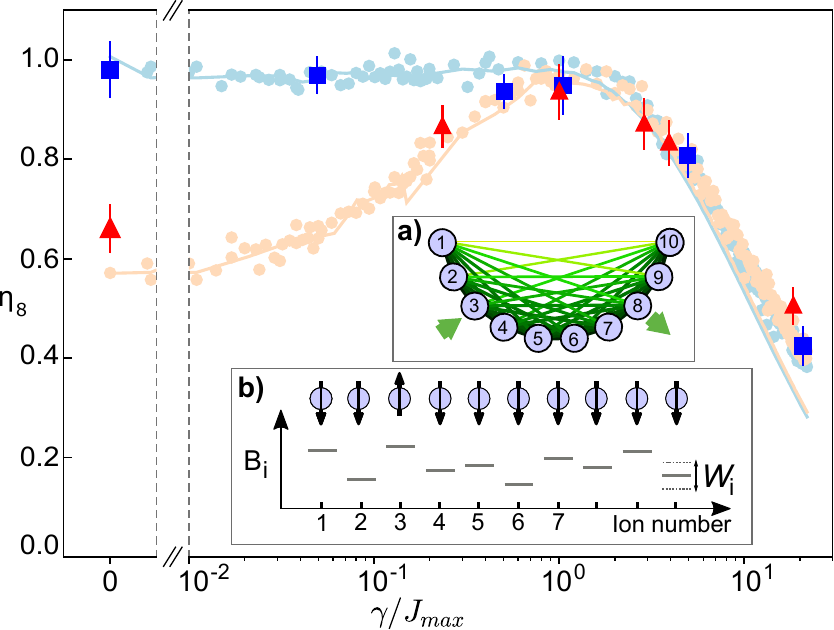}
   \caption{
   \textbf{Main graph:} 
   Transport efficiency $\eta_8$ to the target (ion 8) under different strengths of static disorder (blue: $B_\text{max}=0.5\cdot J_\text{max}$, red: $B_\text{max}=2.5\cdot J_\text{max}$) and Markovian-like dephasing with rate $\gamma$. 
   Experimental points (shown as dark squares and triangles) result from averaging over $20-40$ random realizations of disorder and noise, with $25$ experimental repetitions each. Error bars are derived via  bootstrapping, based on 1000 samples (see~\cite{note:bootstrap} for details). ENAQT is seen to be most advantageous around $\gamma = J_\text{max}$.
   The data agrees well with theoretical simulations of the coin-tossing random process (light bullets) realized in the experiment, while simulations with ideal Markovian white noise (lines) underestimate ENAQT at large $\gamma$. The simulation averages over 300 random realizations.
   \textbf{Inset a):} Sketch of the transport network. The ions experience  a long-range coupling, with darker and thicker connections indicating higher coupling strengths. The green arrows denote the source (3) and the target (8) for the excitation in the ion network.
   \textbf{Inset b):} Sketch of the  ion chain representing interacting spin-$\nicefrac{1}{2}$ particles as blue circles, with the spin states  denoted by black arrows. The ions are subject to random static and dynamic on-site excitation energies, indicated by $B_i$ and $W_i(t)$.
   }
\label{fig:Markov_result}  
\end{figure}

In this work, we study ENAQT in a controlled network of 10 coupled spins subject to static disorder and dephasing noise (see inset (a) in Fig.~\ref{fig:Markov_result}). The network is realized in a system of trapped ions following a recent proposal~\cite{Trautmann:2017}. Our approach enables us to investigate the role of ENAQT in a controlled quantum network that does not have a simple lattice structure restricted to close-neighbor interactions. 
First, applying white dephasing noise of increasing strength, we observe a crossover from coherent dynamics and Anderson localization to ENAQT, and finally a suppression of transport due to the quantum Zeno effect. In the regime where ENAQT is most effective, we find that the transport reveals coherences only at very short times, and that the spread of the excitation is mostly diffusive. Finally, we show that non-Markovian dephasing can maintain coherences longer than white noise, with a strong influence of the structure of the noise spectrum on the transport efficiency.

\emph{Experimental implementation.---}
\label{sec:Experimental Implementation}
The nodes of our quantum network are encoded into (pseudo-) 
spin-\nicefrac{1}{2} particles, represented by two internal electronic states of $^{40}\mathrm{Ca}^{+}$ ions trapped in a linear Paul trap~\cite{Schindler:2013}. We define the state  $\ket{\text{S}_\frac{1}{2}, m=+\frac{1}{2}} $ as spin down $\ket{\downarrow}$ and $\ket{\text{D}_\frac{5}{2}, m=+\frac{5}{2}}$ as spin up $\ket{\uparrow}$. 
A spin--spin interaction Hamiltonian is realized by global laser pulses coupling the electronic states of all ions~\cite{Jurcevic:2014}. In the subspace with a single spin excitation $\ket{\uparrow}$, the Hamiltonian is given by 
\begin{equation}
H_1=\hbar \sum_{i\neq j} J_{ij}(\sigma_i^{+}\sigma_j^{-}+h.c.). 
\label{eq:Hamiltonian0}
\end{equation}
Here, $\sigma_i^{+} (\sigma_i^{-})$ are the spin raising (lowering) operators for site $i$.  This Hamiltonian describes the hopping of spin excitations between sites $i$ and $j$ and conserves the total magnetization, i.e. the number of spins in the excited state $\ket{\uparrow}$ is preserved. The hopping rates follow an approximate power-law, \mbox{$J_{ij} = J_\mathrm{max}/ |i-j|^{\alpha}$}, with peak strength $J_\mathrm{max}$ between $(2\pi)~\SI{28}{Hz}$
and $(2\pi)~\SI{33}{Hz}$ and exponent $\alpha = 1.22$~\cite{Jurcevic:2015}. Inset (a) of Fig.~\ref{fig:Markov_result} depicts this quantum network.

Static disorder, disturbing the quantum network, is represented as on-site excitation energies $\hbar B_i$, see Fig.~\ref{fig:Markov_result} (b) and Eq.~(\ref{eq:Hamiltonian}) below. The values $B_i$ are randomly sampled from a uniform distribution $[-B_\text{max},B_\text{max}]$, with $B_\text{max} \in \{ 0.5, 2.5 \} \cdot J_\mathrm{max}$.  
In the experiment, these disorder energies are realized by laser beams focused to single ions and introducing precisely controlled AC-Stark shifts on the encoded spin states~\cite{Smith:2016}. For this, we apply multiple radio frequencies to an acousto-optical deflector, generating a set of laser beams to simultaneously address multiple ions. 

Moreover, we can temporally modulate the AC-Stark shifts, employing an arbitrary-wave-form generator with a switching time much faster than all other time scales. Using this technique, we are able to engineer time-dependent on-site energies $\hbar W_i(t)$, which induce dephasing between the $\ket{\downarrow}$ and $\ket{\uparrow}$ states. In this way, we simulate a stationary noise process with vanishing mean, $\braket{\braket{W_i(t)}}=0$, and 
broadly tunable spectral power~\cite{Rieke:1999,Miller:2012}
\begin{equation}
\begin{split}
\label{eq:S(w)}
&S(\omega)= \lim\limits_{T \rightarrow \infty} \frac{1}{T} \int_{0}^T  \int_{0}^T \braket{\braket{W_i(t)W_i(t')}} \mathrm{e}^{i\omega (t-t')} dt' dt \,,
\end{split}
\end{equation}
where $\braket{\braket{\bullet}}$ denotes averaging over noise trajectories. Cross talk between neighbouring spins and subharmonics of the driving frequencies are negligible, so noise at different sites is uncorrelated.
Including static disorder and dynamical dephasing noise, the Hamiltonian $H_1$ becomes  
\begin{equation}
H=\hbar \sum_{i\neq j} J_{ij}(\sigma_i^{+}\sigma_j^{-}+h.c.)+
\hbar \sum_i(B_i+W_i(t))\,\sigma_i^z \,.
\label{eq:Hamiltonian}
\end{equation}
where $\sigma^z$ is the Pauli-$z$ matrix. 

To investigate ENAQT, we introduce an excitation at time $t=0$ at the source site $i_\mathrm{source}=3$ (see inset (a) of Fig.~\ref{fig:Markov_result}) by preparing spin  $i_\mathrm{source}$ in the  $\sigma_i^z$ eigenstate $\ket{\uparrow}$ while keeping all other spins in the eigenstate $\ket{\downarrow}$. 
We observe the transport of the excitation through the network to the target site  $i_{\mathrm{target}}=8$ under the Hamiltonian $H$ in Eq.~(\ref{eq:Hamiltonian}), for both Markovian and non-Markovian dephasing. The source and target sites are chosen such that the transport dynamics is not immediately influenced by boundary effects. 
We define the transport efficiency to a particular site $i$ by \mbox{$\eta_i\equiv \int_0^{t_{\mathrm{max}}} \mathrm{d}t p_{{i}}(t)$}. 
Here, \mbox{$p_{i}(t)=(\braket{\sigma_i^z(t)}+1)/2$} is the instantaneous probability to find the excitation at \mbox{site $i$} and \mbox{$t_{\mathrm{max}} = \SI{60}{ms} \approx 11.7 / J_\text{max}$} is the system's evolution time. The time is chosen such that the evolution is long enough to observe ENAQT and short enough to minimize decoherence from amplitude damping due to spontaneous decay~\cite{Trautmann:2017}. 
Any residual amplitude damping effect is eliminated by postselecting measurements with a single excitation in the system. Typically more than 77$\%$ of the measurements lie within this subspace. 

\emph{Markovian dephasing.---}
\label{subsec:Markovian dephasing}
We first study ENAQT in the regime where $W_i(t)$ can be described as white (or Markovian) noise, i.e. $S(\omega)=\text{const.}$ In the experiment, every \mbox{$\Delta T = \SI{100}{\micro\second}$ ($\SI{200}{\micro\second}$)} we randomly sample $W_i$ between \mbox{$\{- \frac{W_\text{max}}{2},  \frac{W_\text{max}}{2}\}$} with equal probabilities  (equivalent to tossing a coin). 
As the \grqq coin tossing rate" \mbox{$\lambda = 1/\Delta T$} is much faster than the maximal hopping $J_\mathrm{max}$, over the relevant frequency range this process is well approximated as white noise \mbox{$S(\omega)=\frac{W_\text{max}^2}{\lambda} = \gamma$} with dephasing rate $\gamma$. 
We apply the dephasing noise to our system under (i) weak static disorder $B_\text{max} = 0.5 \cdot J_\text{max}$ and (ii) strong static disorder $B_\text{max} = 2.5 \cdot J_\text{max}$. 

Figure~\ref{fig:Markov_result} shows the measured transport efficiency $\eta_\text{8}$ as a function of $\gamma/J_\text{max}$: 
Weak static disorder $B_\text{max} < J_\text{max}$ (blue markers) does not affect transport considerably. However, with additional noise at a level beyond $\gamma = J_\text{max}$ the transport efficiency gradually decreases. This regime, where noise is the dominant effect and inhibits quantum transport, is known as the quantum Zeno regime.
Under strong static disorder $B_\text{max} > J_\text{max}$ (red markers), the phenomenology becomes even richer: At weak dephasing, $\gamma < J_\text{max}$,  excitation transport is suppressed corresponding to Anderson localization. Around  $\gamma\approx J_\text{max}$, the noise cancels destructive interference causing the localization and thereby enhances the transport efficiency, which is the hallmark of ENAQT. 
For strong noise, $\gamma > J_\text{max}$, the quantum Zeno effect again suppresses transport. 

The experimental results agree well with theoretical simulations of the coin-tossing random process (light bullets in Fig.~\ref{fig:Markov_result}). At very strong dephasing, the shift $W_\text{max}$ becomes comparable to the coin flipping rate $\lambda$ and the Markovian approximation is no longer fulfilled. In this case, deviations from ideal Markovian white noise (lines) become noticeable, as discussed in~\cite{Trautmann:2017}. 
Such non-Markovian effects will be further discussed later, after analyzing the coherence properties of the quantum transport.

\begin{figure}[t!] 
   \centering
   \includegraphics[scale=1]{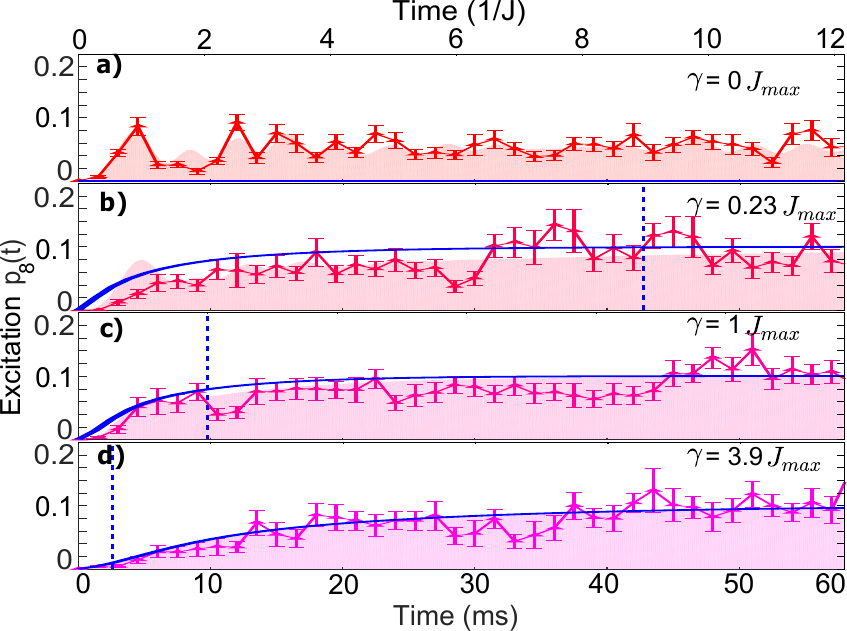} 
   \caption{Excitation probability at ion 8 as a function of time, for strong static disorder $B_\text{max} = 2.5\cdot J_\text{max}$ and increasing dephasing rate, from (a) to (d) $\gamma/J_\text{max} = 0~|~0.23~|~1~|~3.9$. 
Each data set (red to magenta triangles) results from averaging over $20-40$ random realizations, with $25$ repetitions each. Error bars are derived with bootstrapping~\cite{note:bootstrap}, based on 1000 samples. 
With increasing $\gamma$, coherent oscillations damp out and the data converges towards a model following diffusive, classical rate equations (blue solid line). This theoretical approximation is valid for times $t \gg 1/\gamma$. (The crossover $t_c = 1/\gamma$ is illustrated by a blue dashed line.)  
The shaded areas show the time evolution of a theoretical model with ideal Markovian noise, averaged over $100$ random realizations.}
   \label{fig:TimeEvol}
\end{figure}

\emph{Coherent dynamics in excitation transport.---}
The role of coherences in ENAQT has been much discussed in the context of exciton transport in photosynthetic complexes~\cite{Engel2011,Kassal2013,Fassioli2013,RomeroPhotosynthesis2013,Chenu2014}. 
To investigate how coherences affect excitation transport in our system, we observe the time-resolved dynamics of the excitation probability of spin 8, $p_{8}(t)$, for strong static disorder and several levels of dephasing noise (see Fig.~\ref{fig:TimeEvol}). Without dephasing noise, $\gamma = 0$, we find strong oscillations, indicative of quantum coherent transport. Already in the regime where ENAQT becomes relevant, $\gamma \approx J_\text{max}$, however, the noise damps out any perceivable coherent oscillations, with spurious oscillations lying in the range of statistical fluctuations. 
Further, at sufficiently large times, $t \gg 1/\gamma$, and for large values of $\gamma$, the dynamics of the excitation probability of spin $i$, $p_{i}(t)$, is well described by a classical rate equation (blue solid lines, Fig.~\ref{fig:TimeEvol}). Here, the coherences between sites have been adiabatically eliminated (see Supplementary Material), resulting in the equation 
\begin{equation}
\label{eq:dotrhonn}
\dot p_{i} = \sum_{\ell\neq i}\Gamma_{\ell i} (p_{\ell} - p_{i})\,,
\end{equation}
with the classical hopping rate \mbox{$\Gamma_{\ell i} =\frac{\gamma J_{i\ell}^2}{4\left(B_{i}-B_{\ell}\right)^2+\gamma^2}$}, derived from the experimental spin--spin coupling matrix $J_{i\ell}$ and the applied static on-site energies $B_i$ as well as dephasing noise rate $\gamma$. 
This set of coupled differential equations describes a purely diffusive transport of the spin excitation. 
For weak dephasing, we observe deviations from rate equation \eqref{eq:dotrhonn} at short times, which indicates a temporal crossover from ballistic to diffusive transport, similar to what has recently been resolved in classical Brownian motion \cite{Huang2011}. 
With increasing dephasing strength, the observed coherences are damped and the system converges to a diffusive rate equation. This highlights the fact that Anderson localization is a wave phenomenon caused by destructive interference, which is lifted by dephasing. 

\emph{Crossover from ballistic to subdiffusive  transport.---}
We can quantify the transport behaviour by examining the spatial dispersal of the excitation, i.e.\ by measuring the spatial width $\sigma_\text{WP}$ of the excitation wave packet. 
This analysis is analogous to experiments with ultracold atoms in a momentum space lattice \cite{Gadway2017} and to experiments in a photonic system on a discrete quantum walk~\cite{Schreiber2011}. 
We calculate the width via the spread from source spin $i_3$:
\mbox{$ \sigma_\text{WP}(t) \approx  \sqrt{2\left(\sum_{i>i_3}{p_{i}(t)\cdot (i-i_3)^2}\right)}$} (this formula is chosen to reduce boundary effects, see Supplementary Material for details). Depending on the relationship to time, $\sigma_\text{WP}(t) \propto t^C$, one distinguishes between `normal diffusion' as it occurs in classical random walks ($C = 0.5$), `subdiffusion' ($0<C<0.5$), and `superdiffusion' ($C>0.5$). The case $C=1$ is referred to as ballistic transport.
As we now show, we observe ballistic, diffusive, and subdiffusive behavior in our experiment. 

\begin{figure}[t!] 
   \centering
   \includegraphics[scale = 1]{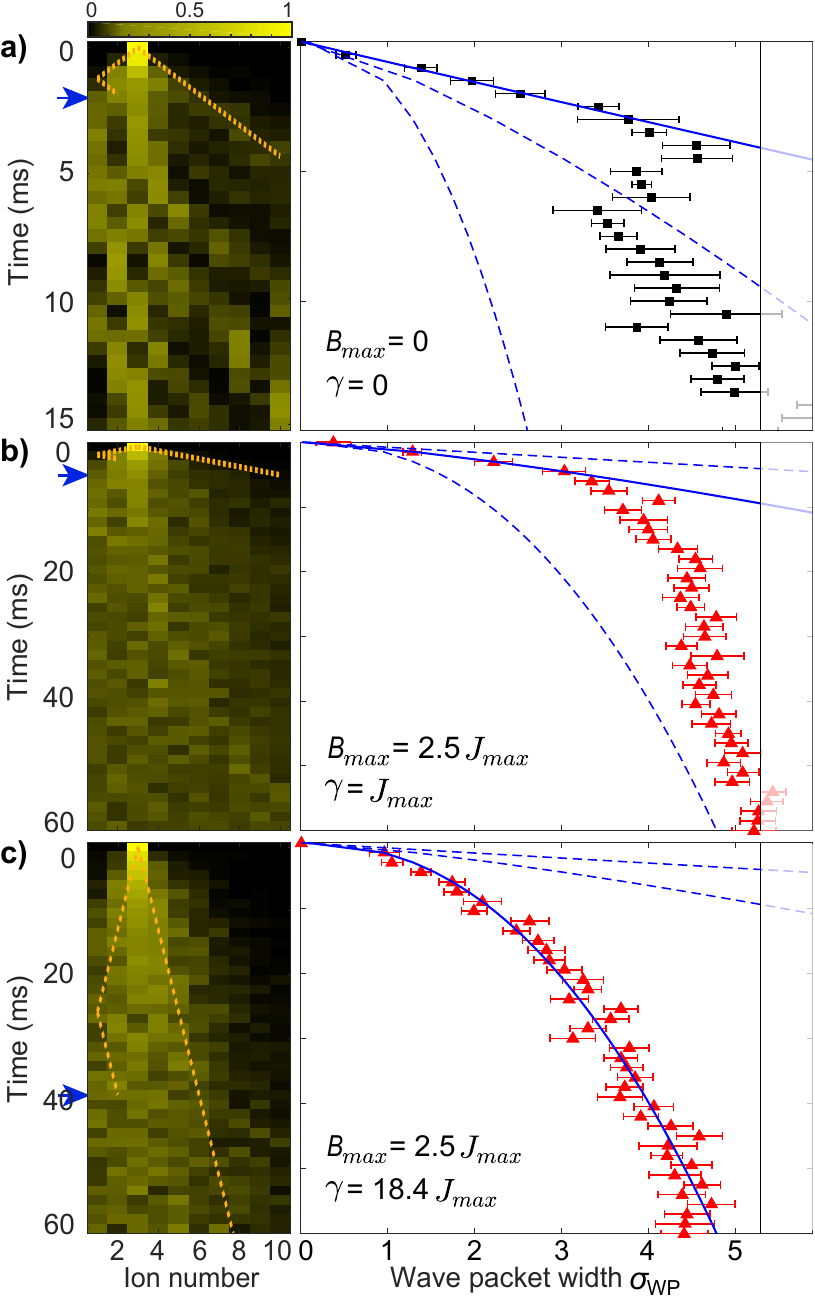} 
   \caption{\textbf{Left panels:} Single-ion resolved excitation dynamics $p_i(t)$ for  \textbf{(a)} the unperturbed system (no static disorder and no noise),  \textbf{(b)} static disorder $B_\text{max} = 2.5 \cdot J_\text{max} $ with dephasing $\gamma = J_\text{max}$, and  \textbf{(c)} static disorder $B_\text{max} = 2.5 \cdot J_\text{max} $ with $\gamma = 18.4 \cdot J_\text{max}$. The orange dotted line shows the maximum speed at which an excitation spreads (see Supplementary Material) in order to estimate until when the reflection from the left boundary can be neglected (blue arrows). 
   \textbf{Right panels:} Spatial width of the excitation wave packet $\sigma_\text{WP}(t)$, calculated from the data in the left panels. Blue solid lines are fits of the form $\sigma_\text{WP} = A \cdot t^C$ (fits from the respective other panels are included as dashed lines for comparison). The vertical black line at $\sigma_\text{max} = 5.3$ is the expected maximum of the wave packet width, for a single excitation distributed equally over all ions.
   \textbf{(a)} %No static disorder and no dephasing.  
   $A = (1.2 \pm 0.8) \cdot 10^{-3},  C = {1.01 \pm 0.09}$
   %for B = 0.5: $A = (1.0 \pm 0.7) \cdot 10^{-3},  C = {1.02 \pm 0.09}$. 
   \textbf{(b)} 
   %Static disorder $B_\text{max} = 2.5 \cdot J_\text{max} $ and dephasing $\gamma = 1 \cdot J_\text{max}$. 
   $A = (5.1 \pm 7.6) \cdot 10^{-3}, C = {0.76 \pm 0.18}$.
   %for B = 0.5: $A = (1.3 \pm 0.3) \cdot 10^{-2}, C = {0.66 \pm 0.03}$.
   \textbf{(c)} 
   %Static disorder $B_\text{max} = 2.5 \cdot J_\text{max} $ and strong  dephasing $\gamma= 18.4 \cdot J_\text{max} $. 
   $A = (3.9\pm 0.9) \cdot 10^{-2}, C = {0.44 \pm 0.02}$. 
   %for B = 0.5: $A = (2.9\pm 0.6) \cdot 10^{-2}, C = {0.46 \pm 0.02}$. 
   All error bars are derived via bootstrapping~\cite{note:bootstrap}, based on 100 samples.} 
   \label{fig:WavePacket}
\end{figure}

The excitation dynamics $p_i(t)$ for three exemplary parameter values is displayed in the left column of Fig.~\ref{fig:WavePacket}. At small $\gamma$ an interference pattern is clearly visible. This hallmark for coherence is rapidly washed out as $\gamma$ increases. 
We fit a power law of the form \mbox{$\sigma_\text{WP}(t) = A \cdot t^C$} to the width of the wave packet (see Fig.~\ref{fig:WavePacket}, right column), only including data up to the time where the excitation has been reflected from the left boundary back to ion 2 (Fig.~\ref{fig:WavePacket}, left column). 
In this way, we exclude data dominated by boundary effects. 
Without any disorder and noise, the width increases linearly in time with $C = {1.01 \pm 0.09}$, corresponding to ballistic spreading. 
In the regime around $\gamma = J_\text{max}$ (Fig.~\ref{fig:TimeEvol} (b)), where ENAQT is most efficient, we find that within very short times $t \sim 1/J_\text{max}$ the transport evolves from ballistic to mainly diffusive dynamics (as theoretically predicted in Ref.~\cite{Knap2017}), yielding $C = {0.76 \pm 0.18}$. 
For strong dephasing, $\gamma=18.4\cdot J_\text{max}$, we observe subdiffusive transport with a power exponent $C = {0.44 \pm 0.02}$. Based on theoretical simulations, we conclude that subdiffusive dynamics is caused by the long-range interactions in our system. 

\emph{Non-Markovian dephasing.---}
\label{subsec:Non-Markovian dephasing}
In Fig.~\ref{fig:Markov_result}, the experimentally observed transport efficiencies for $\gamma > J_\text{max}$ are higher than the simulated values for ideal Markovian noise. 
This discrepancy could indicate that non-Markovian effects can increase the transport efficiency. To investigate non-Markovian dephasing further, we study ENAQT under noise with a spectral density function $S(\omega)$ of Lorentzian shape, which we generate using the frequency-domain algorithm described in Ref.~\cite{Percival92simulatinggaussian}. We choose a single random configuration of static disorder ($B_\text{max} = 2.5 \cdot J_\text{max}$) in order to have full knowledge of the disordered system and its eigenvalues. 

\begin{figure}[t!] 
   \centering
   \includegraphics[scale = 1]{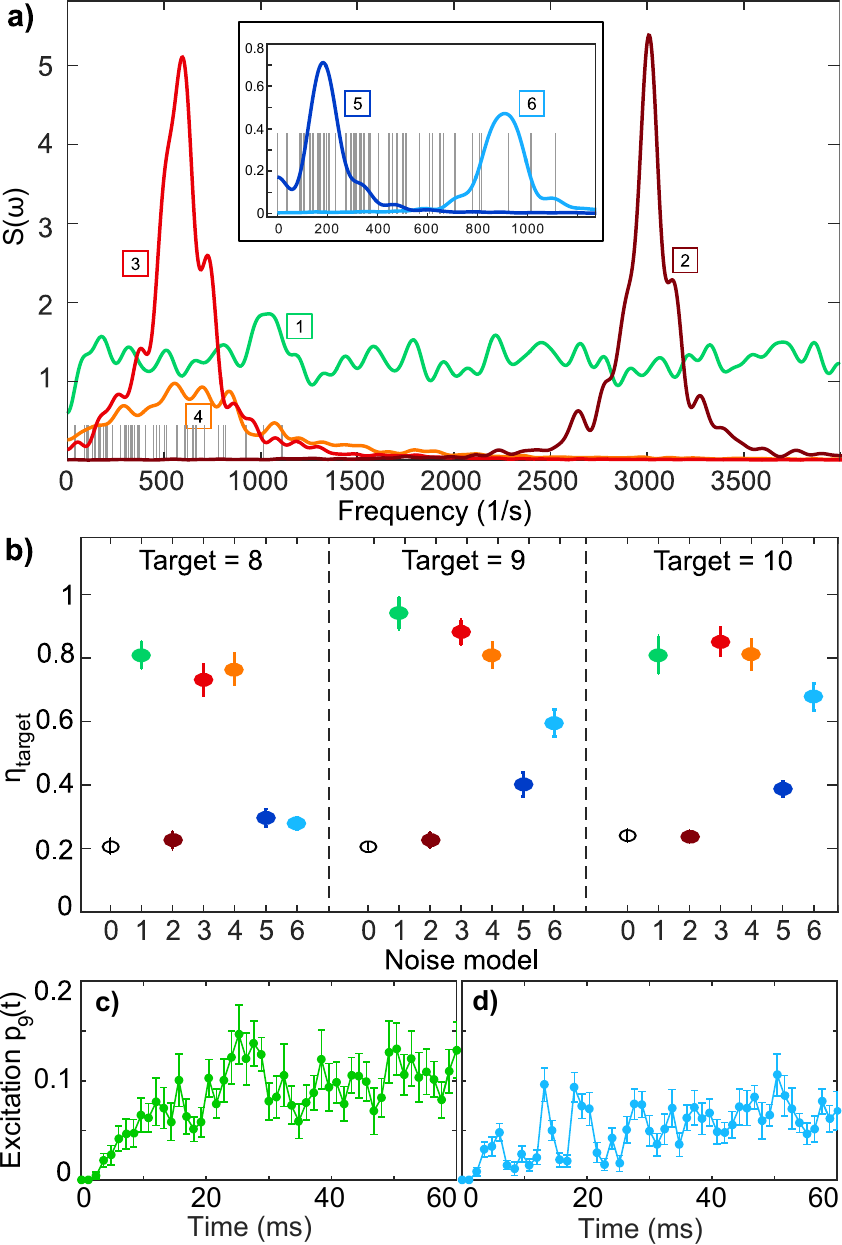} 
   \caption{Excitation transport efficiency $\eta$ under strong static disorder and different noise  models. 
   \textbf{a)} Spectral density functions $S(\omega)$ of the applied noise models. (1) white noise, (2)-(6) non-Markovian noise models of Lorentzian shape. 
   The curves are averaged over $\sim 30$ random realizations, each generated by a Gaussian random process based on $600$ sampling points. The inset shows a zoom into the low-frequency domain. 
 Vertical grey lines denote the difference frequencies between all eigenenergies of the disordered system.
\textbf{b)} Comparison of $\eta_{\mathrm{target}}$ to target ion 8, 9, and 10 for the noise models shown in (a), as indicated by corresponding colors and numbering. The black circle shows suppressed efficiency under strong static disorder without any noise. 
While broadband noise in the correct frequency range generically enhances transport efficiencies, for narrowband noise the enhancement depends on the source and target ions. 
 Each data point results from averaging over $25-30$ random realizations, with $15$ repetitions each. Error bars are derived with bootstrapping~\cite{note:bootstrap}, based on 1000 samples.
 \textbf{c)-d)} Excitation probability of target ion 9 as a function of time under strong static disorder $B_\text{max} = 2.5\cdot J_\text{max}$. Panel c) shows the result for the Markovian noise model (1). Oscillations, indicating coherent dynamics, are strongly damped. Panel d) shows the effects of a narrowband Lorentzian noise model covering only a few eigenstates (model 6). Here, coherences are maintained stronger and are clearly discriminable from measurement errors up to $\sim$\SI{30}{ms}.}
   \label{fig:NonMarkovian_results}
\end{figure}

From Fig.~\ref{fig:NonMarkovian_results}, we see that the spectral structure of the noise model has a strong influence on the transport efficiency: 
Non-Markovian structured noise that covers all difference frequencies of the spin system's eigenenergies (models 3 and 4) can enhance excitation transport as much as white noise (model 1), and the efficiency is similar  for different target ions (Fig.~\ref{fig:NonMarkovian_results} (b)). 
Narrowband noise models, instead, only couple a few eigenstates, so the spectral position determines for which target ions excitation transport is enhanced (cf.\ models 5 and 6 in Fig.~\ref{fig:NonMarkovian_results}). 
Integrating the applied local energy shifts over the entire interaction time, we find that with narrowband non-Markovian noise we can achieve similar transport efficiencies as with Markovian noise, but already at half the energy cost (cf.\ noise models 1 and 6 in Fig.~\ref{fig:NonMarkovian_results} (a)). 
Further, panels (c) and (d) in Fig.~\ref{fig:NonMarkovian_results} show that coherences can be maintained better for narrowband noise models than for Markovian-like noise.

\emph{Conclusion.---}
We have experimentally analyzed a quantum network under static disorder and dynamic noise, realized in a string of 10 trapped ions. We observed effects of Anderson localization in the absence of noise, an increased transport efficiency by ENAQT at intermediate noise levels, and finally suppression of quantum transport under strong noise due to the quantum Zeno effect. 
Further, we have found that coherences play a role only in the localized regime (at very low noise strengths) or at very short times. In all other regimes of Markovian noise, the dynamics is well captured through a diffusive rate equation describing a classical random walk. 
Finally, we found that the structure of non-Markovian dephasing strongly influences quantum transport, with the possibility to reach as large efficiencies as with white noise while maintaining long-lived coherences. 

In the future, it will be interesting to study the possibility of stochastically accelerated hyper-transport, generated, e.g., by time-evolving disorder \cite{Levi2012}. 
Further, our approach allows one to investigate quantum transport with multiple interacting excitations or to study localization using out-of-time-ordered correlators  (OTOCs)~\cite{OTOCs_Localized,OTOCs_Ising}. 

\section{Supplementary Material}

Here, we discuss the derivation of the classical rate equation used to describe the diffusive part of the system dynamics. Furthermore, a useful way for calculating the wave packet width with reduced boundary effects is shown. 

\subsection{Classical rate equation from dephasing noise}
The main idea for arriving at a classical rate equation is to adiabatically eliminate the coherences between sites, an approximation which becomes valid at times larger than the inverse dephasing rate. We start from the full master equation for our system, which reads  
\begin{equation}
\dot \rho = -\mathrm{i} [H,\rho] +\mathcal{L}\rho\,,
\end{equation}
with the Lindblad superoperator for dephasing noise $\mathcal{L}X=\sum_i \frac{\gamma_i}{2} (2\sigma_i^+\sigma_i^- X \sigma_i^+ \sigma_i^- - \sigma_i^+\sigma_i^- X - X \sigma_i^+ \sigma_i^- )$.  

In the single-excitation subspace and in the case where noise and disorder dominate over the hopping terms, it is convenient to work in the basis spanned by the states $\ket{i}=\sigma_i^+\ket{\Downarrow}$, with $i=1\dots N$ and $\ket{\Downarrow}$ the fully polarized state. In this basis, the excitation probabilities (`populations') evolve as 
\begin{equation}
\dot \rho_{ii} = -\mathrm{i} \sum_{\ell\neq i}(H_{i\ell}\rho_{\ell i}-\rho_{i\ell}H_{\ell i})\,,
\end{equation}
where we define $\rho_{ij}=\bra{i}\rho\ket{j}$ and analogously for $H_{ij}=\bra{i}H\ket{j}$
The coherences for $i\neq j$ evolve as 
\begin{eqnarray}
\label{eq:dotrhonm}
\dot \rho_{ij} &=& -\mathrm{i} \left(\sum_{\ell\neq i}H_{i\ell}\rho_{\ell j}-\sum_{\ell\neq j}\rho_{i\ell}H_{\ell j}\right)\nonumber\\
&+&\left[-\mathrm{i} (H_{ii}-H_{jj})-\frac{\gamma_i+\gamma_j}{2}\right]\rho_{i j}\,. 
\end{eqnarray}
Here, the terms $H_{i\ell}=J_{i\ell}$ ($i\neq\ell$) describe the hoppings and $H_{ii}=2 B_i$ the on-site disorder (up to a constant). 

Under the assumption that the diagonal terms $H_{ii}$ and $\gamma_i$ are the dominating energy scales, we can adiabatically eliminate the coherences. Formally, this amounts to setting their time-derivatives  to zero, which becomes valid on the ``slow" time scales on which the populations evolve, $t\gg 1/\left|\mathrm{i} (H_{ii}-H_{jj})+\frac{\gamma_i+\gamma_j}{2}\right|$. 

Solving the Eq.~\eqref{eq:dotrhonm} for $\dot \rho_{ij}=0$ to leading order in the hoppings, i.e., assuming $H_{ij}\ll\left|\mathrm{i} (H_{ii}-H_{jj})+\frac{\gamma_i+\gamma_j}{2}\right|$, we obtain $\rho_{ij}=\frac{H_{ij}(\rho_{ii}-\rho_{jj})}{H_{ii}-H_{jj}-\frac{\gamma_i+\gamma_j}{2}}$. Inserting this expression into Eq.~\eqref{eq:dotrhonn}, we obtain the result
\begin{equation}
\dot \rho_{ii} = \sum_{\ell\neq i}\Gamma_{\ell i} (\rho_{\ell \ell} - \rho_{ii})\,,
\end{equation}
with 
$\Gamma_{\ell i} =\frac{\frac{\gamma_i+\gamma_\ell}{2}H_{i\ell}H_{\ell i}}{\left(H_{ii}-H_{\ell \ell}\right)^2+\left(\frac{\gamma_i+\gamma_\ell}{2}\right)^2}$. 
By setting $\gamma_i=\gamma\,\,\,\forall\, i$, this set of coupled differential equations describes the diffusive evolution of the populations $p_i=\rho_{ii}$ according to the classical rate equation \eqref{eq:dotrhonn} given in the main text. 

\subsection{Estimation of the wave packet width}
The left panels in Fig.~\ref{fig:WavePacket} show single-ion resolved excitation dynamics $p_i(t)$, which is used to calculate the spatial width of the excitation wave packet $\sigma_\text{WP}(t)$ over ion 3 to 10. We start from the common definition of the wave packet width 
$ \sigma_\text{WP}(t) = \sqrt{\braket{\hat{x}^2}-\braket{\hat{x}}^2} = \sqrt{\left(\sum_{i=1}^{10}{p_i(t)\cdot i^2}\right) - \left(\sum_{i=1}^{10}{p_i(t)\cdot i}\right)^2 }$
and rewrite the expression relative to the source site $i_0$:
$ \sigma_\text{WP}(t) =\sqrt{\left(\sum_{i}{p_i(t)\cdot (i-i_0)^2}\right) -  \left(\sum_{i}{p_i(t)\cdot (i-i_0)}\right)^2 }$. 
Since the excitation is inserted off-center, we can increase the spatial and temporal range over which the width is evaluated. For this, we discard the data between source and the nearer boundary, where boundary effects appear early. Instead, we only consider the region between source and the boundary that is farther away. We mirror this region around $i_0$, thus obtaining an imagined system where the excitation spreads symmetrically around $i_0$. This description is valid as long as the boundary effects from the nearer boundary do not influence the data in the evaluated region. 
Mathematically, we split the sums at $i_0$ and assume mirror symmetry, which yields  
$ \sigma_\text{WP}(t) = \sqrt{\splitfrac{\left(\sum_{i<i_0}{p_i(t)\cdot (i-i_0)^2}\right) + \left(\sum_{i>i_0}{p_i(t)\cdot (i-i_0)^2}\right)}{
- \left(\sum_{i<i_0}{p_i(t)\cdot (i-i_0) + \sum_{i>i_0}{p_i(t)\cdot (i-i_0)}}\right)^2 }} \\
\approx  \sqrt{2\left(\sum_{i>i_0}{p_{i}(t)\cdot (i-i_0)^2}\right)}
$. 

A quantitative description of the transport behaviour is gained by fitting a power law $\sigma_\text{WP}(t)=A\cdot t^C$. In order to exclude data in which boundary effects from the nearer boundary become relevant, we fit the data only up to the time where the excitation has hopped from ion 3 to the left boundary and back to ion 2. We estimate this time through a modified hopping strength $\widetilde{J}_{ij}$, consisting of the original hopping rate $J_{ij}$, reduced by the applied  disorder $B_\text{max}$ and dephasing $\gamma$, 
\begin{equation}
J_\text{eff.} = \text{min}\{\widetilde{J}_{ij}, J_{ij}\} \quad , \text{with} \quad \widetilde{J}_{ij} =
\frac{J_{ij}^2}{B_\text{max}^2+\gamma^2}    
\end{equation}
Based on the hopping rate $J_\text{eff.}$, we calculate the maximum speed at which an excitation spreads in our system (see~\cite{Jurcevic:2014} and methods in~\cite{Lanyon2017}) and visualize it as orange dotted light-like cones in the left panels of Fig.~\ref{fig:WavePacket}. 
Since we do not have finite-range interactions, these are not strict maximum speeds. Still, they provide a practically useful description of the excitation spreading in our system.

\clearpage

\end{document}